\documentclass{aa}
\usepackage{natbib,graphicx,hyperref} 
\usepackage[varg]{txfonts}
\newcommand{\rsun}{R_{\odot}}

\newcommand{\RB}{R_{\rm b}}
\begin{document}

\title{Observing and modeling the poloidal and toroidal fields of the  solar dynamo}

\author{R.~H. Cameron\inst{1} \and T.~L Duvall~Jr.\inst{1} \and M. Sch\"ussler\inst{1}
  \and H. Schunker\inst{1}}

\institute{Max-Planck-Institut f\"ur Sonnensystemforschung, 
  G{\"o}ttingen, Germany}
\date{Received ; accepted}

\abstract
{The solar dynamo consists of a process that converts poloidal field to toroidal
  field followed by a process which creates new poloidal field from the toroidal
  field.}
{Our aim is to observe the poloidal and toroidal fields relevant to the global solar
  dynamo and see if their evolution is captured by a Babcock-Leighton dynamo.}
{We use synoptic maps of the surface radial field from the  KPNSO/VT and SOLIS
  observatories to construct the poloidal field as a function of time and latitude,
  and Wilcox Solar Observatory and SOHO/MDI full disk images to infer the longitudinally
  averaged surface azimuthal field. We show that the latter is consistent with an estimate of
  that due to flux emergence and therefore closely related to the subsurface toroidal field.}
{We present maps of the poloidal and toroidal magnetic field of the global solar
  dynamo. The longitude-averaged azimuthal field observed at the surface results from
  flux emergence. At high latitudes this component follows the radial component of the polar
  fields with a short time lag (1-3 years). The lag increases at lower latitudes.
  The observed evolution of the poloidal and
  toroidal magnetic fields is described by the (updated) Babcock-Leighton
  dynamo model.}
{}
\authorrunning{Cameron et al.}
\titlerunning{The Sun's Poloidal and Toroidal fields}
\maketitle

\section{Introduction}
The basic periodicity of the Sun's global dynamo is about 22 years \citep{1919ApJ....49..153H},
consisting of two approximately 11 year activity cycles \citep{1849AN.....28..302S}.
Typically the first sunspots of an  activity cycle appear before solar minimum
(the traditional start time of a cycle) at latitudes of about $\pm 35^{\circ}$.
The latitude at which they subsequently appear decreases as the cycle progresses until they reach about
$\pm 8^{\circ}$ at the
end of the activity cycle \citep{1858MNRAS..19....1C, 1879AN.....96...23S}.
The appearance of sunspots marks the emergence of magnetic flux in the range of
$5 \times 10^{21}$~Mx to $3\times 10^{22}$~Mx through the solar surface \citep{Schrijver:Zwaan:2000}.
The flux emerging in active regions (with fluxes from $1\times 10^{20}$ to $3\times 10^{22}$~Mx)
mostly obeys Hale's law, indicating that it comes from subsurface toroidal field that switches
orientation from one cycle to the next. 

Ephemeral regions have fluxes in the range of  $3\times 10^{18}$ to  $1\times 10^{20}$~Mx. They have a
statistical tendency to obey Hale's law, and begin emerging before the first sunspots of a cycle
at higher latitudes \citep{1979SoPh...64...93M, 1992ASPC...27..335H}.
The earlier high-latitude emergences form part of the extended activity cycle \citep{1988Natur.333..748W},
which groups the activity at different latitudes associated with one cycle. The extended activity cycle
is revealed in several different ways: in coronal bright points and green-line emission
\citep{1988sscd.conf..414A, 2014ApJ...792...12M}; in chromospheric plage \citep{1992ASPC...27..335H}; in the emergence
of bipoles  preferentially obeying Hale's law \citep{1979SoPh...64...93M}; in the longitudinally averaged
azimuthal inclination of photospheric magnetic flux \citep{1974SoPh...39..275H, 1994SoPh..153..131S, 2010ASPC..428..109L};
in the longitudinally averaged photospheric radial \citep{Hathaway2015} and azimuthal field
\citep{1979SoPh...61..233D, 2005ApJ...620L.123U, 2010ASPC..428..109L}.

\section{Observations of the radial and azimuthal fields}

\begin{figure}
\begin{center}
\includegraphics[scale=0.7]{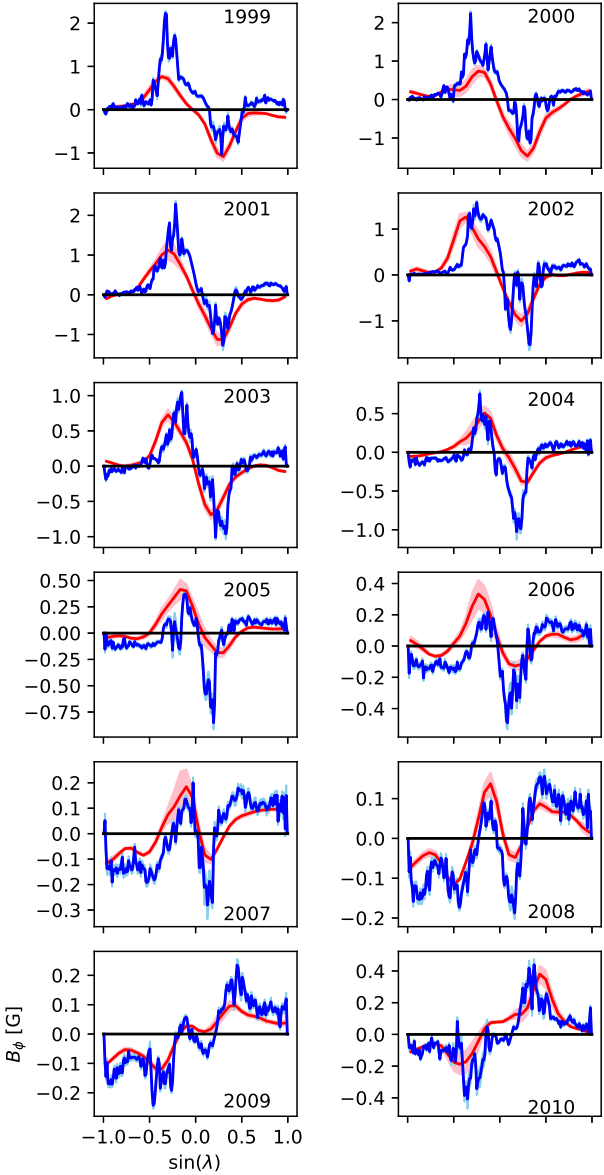}
\caption{Surface longitudinally averaged azimuthal field determined from MDI (blue) and Wilcox Solar Observatory (red).
The pink and light blue shadings show the formal error bars of the fittings. }
\label{fig:cuts}
\end{center}
\end{figure}

We determined the longitudinally averaged surface azimuthal field using the technique described
in \cite{1978PhDT.........4D} and \cite{1979SoPh...61..233D} applied to WSO and SOHO/MDI observations. In brief, the
technique first creates a 'very-deep' line-of-sight magnetogram consisting of a yearly average of the line-of-sight
magnetic field on the solar disk (for the SOHO/MDI data, with each day being first remapped to account for B-angle variations).
This average magnetogram is split into a component that is symmetric with respect to the solar meridian, and a
component that is anti-symmetric. Assuming that the position of the Earth does not affect the location of flux emergence,
the anti-symmetric component at any latitude is proportional to the yearly and longitudinally averaged azimuthal field,
shown in Figure~\ref{fig:cuts}. Some differences between the results from the WSO and SOHO/MDI observations are expected
because the two instruments have different spatial resolutions; for example, the WSO magnetograms do not resolve pores and spots.

During the years near solar maximum (2000 to 2003), the inferred surface azimuthally averaged azimuthal field in the
active latitudes is between 1~G and 2~G. Around this time the inferred surface azimuthally averaged azimuthal field at high latitudes
reverses and then grows to about 0.1~G, which the overall level from 2007 to 2009, when the solar activity was low. The high latitude
azimuthal field after 2003 has the opposite sign in each hemisphere to the field in the low latitudes during maximum.

The observable that corresponds to the relevant poloidal field of the global dynamo is the longitudinally
averaged radial field at the solar surface \citep{2015Sci...347.1333C}, for which we have four activity
cycles of observations, presented in the form of a time-latitude magnetic butterfly map
\citep{Hathaway2015}.

\section{The observed surface toroidal flux as the result of flux emergence}

While the observations are of the surface field, the relevant toroidal field for the solar dynamo is the subsurface
longitudinally averaged azimuthal field, which we can only detect when it emerges through the surface.
Emergences occur in the form of active regions at low latitudes during solar maximum, and in the form of
ephemeral regions  over the entire Sun and throughout the cycle
\citep{1979SoPh...64...93M, 1992ASPC...27..335H}. Both active regions and ephemeral regions have a statistical
tendency to follow Hale's law (i.e. are preferentially aligned in the East-West direction with the orientation
corresponding to a particular activity cycle in each hemisphere). This means that more flux emerges
with an East-West alignment corresponding to Hale's law than disobeying Hale's law. The difference between
these two fluxes is a proxy for the underlying (axisymmetric) toroidal field. 
The preferential East-West alignment of the emerging flux in ephemeral regions shows up in the
longitudinally-averaged azimuthal field at the solar surface, which is directly observable as discussed
in the previous section.

\begin{figure}
\begin{center}
\includegraphics[scale=1.4]{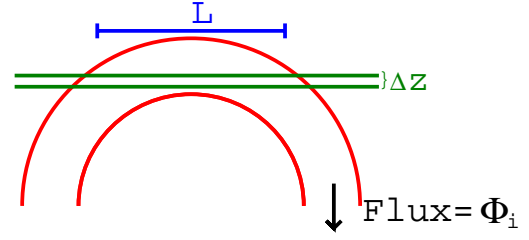}
\caption{Simplified geometry of a flux emergence event. The tube (shown in red) has flux $\Phi_i$.
  The spatial extent of the emergence, the separation of the footpoints, is $L$. During emergence, the
  tube crosses the photospheric layer $\Delta z$ (shown in green) where the polarimetric magnetograph signal is formed.}
\label{fig:ill}
\end{center}
\end{figure}

To quantitatively estimate the average surface azimuthal
field due to flux emergence, we  use the simplified geometry shown in Fig.~\ref{fig:ill}.

We are interested in the average over some area of the solar surface $A$ and time $T$. 
Consider a single flux emergence event: a flux loop with azimuthal flux $\Phi_i$ that emerges
over a length $L_i$ with tilt angle $\gamma_i$ with respect to the azimuthal (EW) direction. The
flux traverses the Zeeman-sensitive layer in the photosphere with an average radial speed $v_{r,i}$ and the observable
azimuthal field, $B_{\phi,i}(\lambda, \phi, t)$, is only non-zero across the emerging region and during
the emergence time, $t_e$, during which the full amount of the flux, $\Phi_i$, traverses the Zeeman sensitive layer of the
photosphere. Integrating over area $A$ and time $T$ we therefore have
    
\begin{eqnarray}
  \Phi_i L_i \cos \gamma_i&=&\iint_{A} \int_T B_{\phi,i}(\lambda,\phi,t) v_{r,i} R_{\odot}^2 cos\lambda
               \mathrm{d}t \mathrm{d}\lambda \mathrm{d}\phi \nonumber \\
  &=& \langle B_{\phi,i}\rangle v_{r,i} T A
\end{eqnarray}
where $\langle B_{\phi,i}\rangle $ is the contribution of the emergence event to the average azimuthal field. 
  The total average azimuthal field, $\langle B_{\phi,i}\rangle$ is then given by the sum over all the emergence events
  across $A$ and during time T, so
\begin{eqnarray}
  \langle B_{\phi}\rangle_{AT}=\sum_i{\langle B_{\phi,i}\rangle}=\frac{1}{AT} \sum_i{\frac{\Phi_i L_i \cos \gamma_i}{v_{r,i}}}  
\end{eqnarray}
where $\langle ... \rangle_{AT}$ indicates the average over area A and time interval T.

In the case of ephemeral regions emerging in the quiet-Sun, we use the results from
\cite{2001ApJ...555..448H}: the total amount of unsigned radial flux emerging in the form of ephemeral
regions over the entire solar surface is about $5\times10^{23}$~Mx/day ($1.8\times10^{26}$~Mx/year),
their spatial extent, $L$, is
about 9~Mm, and about 60\% of the ephemeral regions obey Hale's law. This means 40 percent disobey Hale's law, 
compared to the 60\% which obey Hale's law. The resulting net flux preferentially obeying Hale's law
is hence 20\% of the the total emerging in the form of ephemeral regions, and the E-W aligned component of
this will be 63\% of this (assuming the regions obeying Hale's law emerge with tilts with respect to the equator
uniformly distributed in the range $-90^{\circ}$ to $+90^{\circ}$. 
The rise velocity, $v_r$, is on the order of about 1~km/s \citep[e.g.][]{2012ApJ...745..160G},
resulting in $\langle{B_{\phi}}\rangle=0.11$~G (from Eq.~1), where we have taken the average over
$T=1$~year and area $A$ being the entire solar surface.

Considering the contribution from active regions emerging during maxima, we restrict the area $A$
to the latitudinal extent of both butterfly wings \citep[each of which has a latitudinal extent of about
  $16^{\circ}$][]{2016A&A...591A..46C}, and again consider a yearly averaging. The rate at which flux emerges in the form of 
active regions is $2.3\times 10^{24}$~Mx/year during maximum of cycle 21 \citep{1994SoPh..150....1S}.
The rise velocity, $v_r$, of the toroidal flux through the photosphere in active regions can be estimated from the work of
\cite{2012ApJ...759...72C}. In that paper a range of radial velocities advecting the horizontal magnetic field across
the photosphere in an emerging active region are reported. These include both radially outwards and inwards motions.
We adopt  $v_r=100$~m/s as the characteristic rise velocity,  $L=40$~Mm for a typical active region size, $\cos \gamma =1$
since there is less scatter in the tilt angle of large active regions then for ephemeral regions and $T=1$~year.  
Applying Equation~1 results in a longitudinally averaged azimuthal field value of 1.7~G over
the activity belts.

\begin{figure*}
\begin{center}
\includegraphics[scale=0.4]{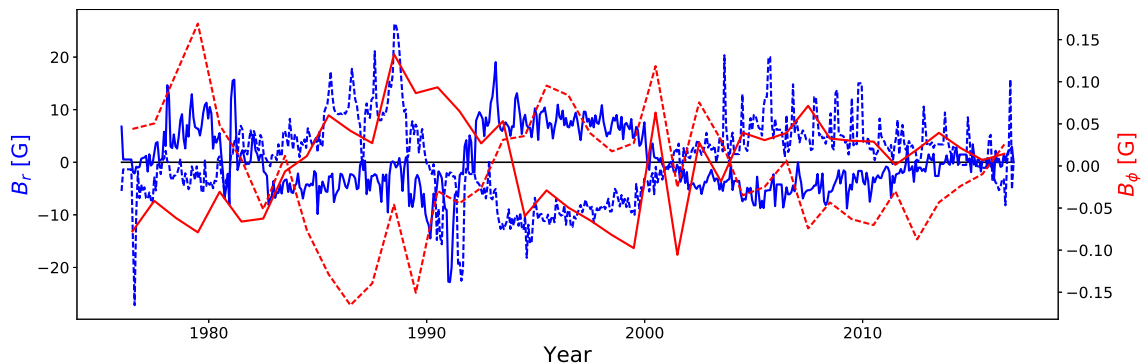}
\caption{The high latitude ($\lambda >50^{\circ}$) surface longitudinally-averaged azithmuthal field from WS0 (red) and
polar ($\lambda >80^{\circ}$) surface longitudinally-averaged radial  field  from KPNSO/VT (blue).
The northern (southern) polar data are shown with the solid (dashed) curves.}
\label{fig:polar}
\end{center}
\end{figure*}

We therefore find that the observed azimuthal field is quantitatively consistent with what we expect from flux emergence. In the quiet Sun,
where flux emergence in the form of ephemeral regions dominates (e.g. most of the Sun during 2008),
the observed average azimuthal field is on the order of 0.1 G (see Fig.~\ref{fig:cuts}) which is comparable to the estimate
for the azimuthal field due to ephemeral regions. In the active Sun, where emergence in the form of active
regions dominates, the observed field is on the order of 1G which is consistent with the expected value from flux emergence
in the form of active regions.

Since flux emergence necessarily depends on the underlying toroidal field, the axisymmetric component of the azimuthal
field determined at the surface in Figure~\ref{fig:cuts} represents a proxy of the subsurface toroidal field distribution 
of the global solar dynamo.

At high latitudes, the axisymmetric component of the azimuthal field follows the axisymmetric component of the radial
component of the polar fields with a slight lag (1-3 years), as can be seen in Figure~\ref{fig:polar}. The toroidal field at lower latitudes
then follow at increasing time lags. This is very highly suggestive of a Babcock-Leighton flux transport dynamo where the field
threading the poles is wound up by differential rotation, and the resulting toroidal flux is advected equatorwards.
The advection of toroidal field away from the pole means
that the toroidal field near the pole follows the polar fields with only a short delay, whereas at low latitudes the
toroidal field of the new cycle must first replace the toroidal field which has already been transported equatorwards
due to advection from higher latitudes.

\section{Consistency with the Babcock-Leighton dynamo}
\begin{figure*}
\begin{center}
\includegraphics[scale=0.7]{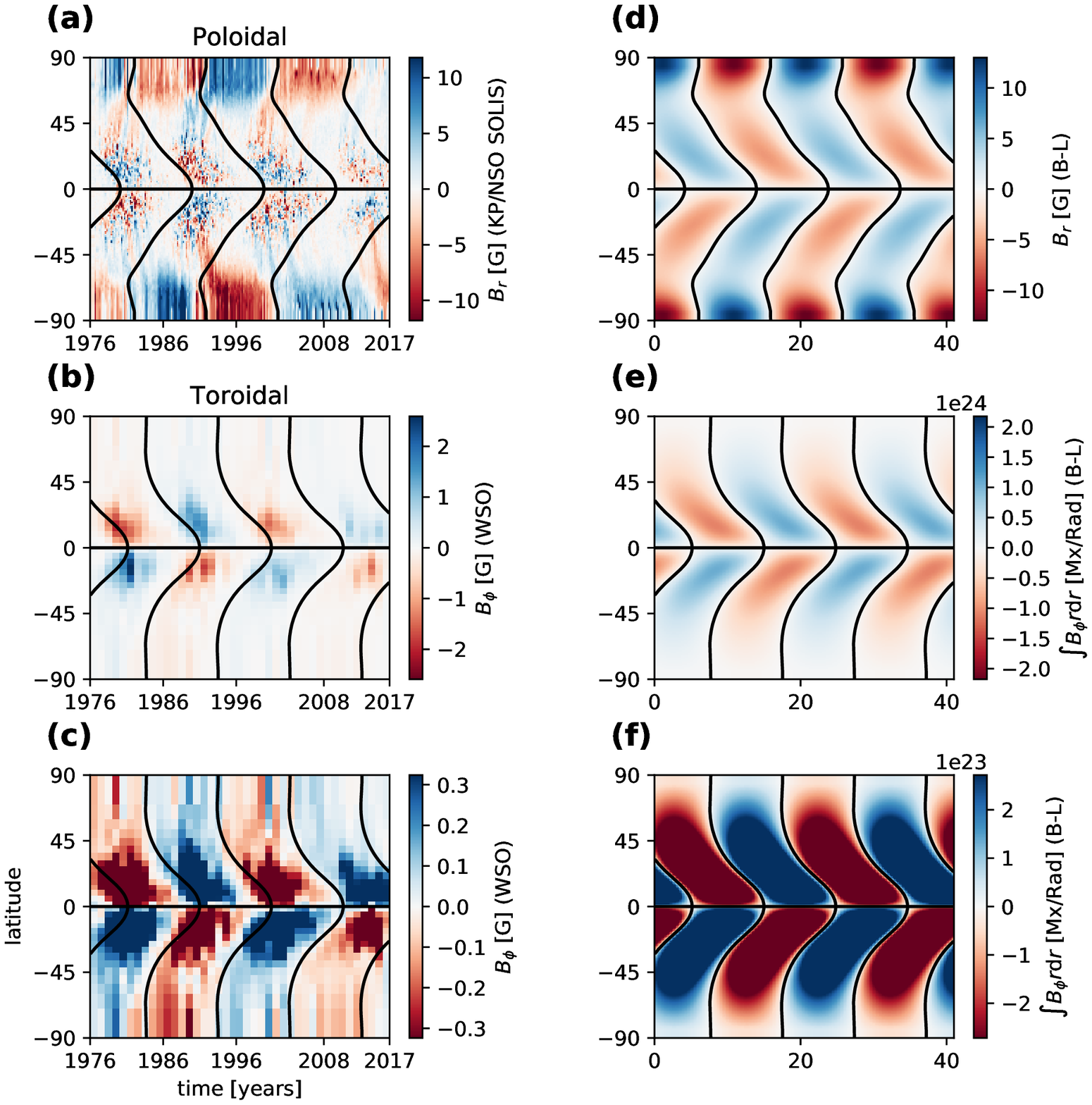}
\caption{Time-latitude diagrams of the observed (left) and simulated (right) magnetic fields.
  Panel (a) shows the observed surface longitudinally averaged radial field, based on KPNSO/VT and SOLIS data. Panel~(b)
  shows the  observed surface longitudinally averaged toroidal magnetic field based on WSO observations,
  and panel (c) is the same as Panel~(b) except saturated to bring out the weaker high-latitude fields.
  Panel~(d) shows the surface radial field from the updated Babcock-Leighton model \citep{2017A&A...599A..52C} and
  Panel~(e) shows the subsurface toroidal flux density from the Babcock-Leighton model, and
  Panel~(f) shows a saturated version of Panel~(e).
  The black lines in all panels show the zero contours from the model. }
\label{fig:sun}
\end{center}
\end{figure*}

A useful way of presenting the time and latitude dependence of the azimuthally averaged observed radial and azimuthal magnetic fields
is in the form of magnetic butterfly diagrams, as shown in Fig.~\ref{fig:sun}.

Given that the observational data covers four cycles, we may now ask whether the
observed evolution is captured by a Babcock-Leighton type dynamo.
For this purpose we use the updated formulation of the model as presented in \cite{2017A&A...599A..52C}.
The model consists of two 1-D evolution equations. The first describes the surface evolution of
the poloidal field,
\begin{eqnarray}
  \frac{\partial a}{\partial t}&=&
  -\frac{U_{\theta,{R_\odot}}}{R_\odot\sin\theta} 
   \frac{\partial (a \sin\theta)}{\partial \theta}\nonumber \\
  & & +\frac{\eta_{R_\odot}}{R_\odot^2} \frac{\partial}{\partial \theta}
  \left(\frac{1}{\sin\theta}\frac{\partial
  (a\sin\theta)}{\partial\theta} 
  \right) + a_{\rm S}(\theta,t)\,,
\label{eqn:a}
\end{eqnarray}
where
\begin{equation}
a(\theta,t)=\frac{1}{\sin\theta}\int_{0}^{\theta} \sin\theta' \, R_\odot^2 B_{r,R_\odot} \mathrm{d}\theta'.
\label{eq:def_a}
\end{equation}
$U_{\theta,{R_\odot}}$ is the surface meridional flow, $\eta_{R_\odot}$ is the turbulent diffusion
coefficient for the dispersion of radial field at the surface in the latitudinal direction,
and $a_{\rm s}$ is the source term which corresponds to flux emergence of active regions preferentially
obeying Joy's law (discussed below). This is the azimuthally averaged form of the surface flux transport model
\citep{2014SSRv..186..491J}.
The second evolutionary equation governs the evolution of the
subsurface toroidal flux (per radian in latitude), 
$b(\theta,t) = \int_{\RB}^{R_{\odot}}B_{\phi}(r,\theta) r \mathrm{d}r$,
\begin{eqnarray}
\frac{\partial b}{\partial t}
  &=& \sin\theta\,
  R_\odot^2 B_{r,{R_{\odot}}} \epsilon \left( \Omega_{R_\odot} -
  \Omega_{R_{\rm NSSL}} \right) \nonumber \\
 & &  -\frac{\partial \Omega_{R_{\rm NSSL}}}{\partial \theta}
    \int_0^{\theta} \sin\theta\, R_{\odot}^2 B_{r,{R_{\odot}}}
  \mathrm{d}\theta   \nonumber \\ 
  & & -\frac{1}{\rsun} \frac{\partial}{\partial\theta} 
      \left( V_\theta \, b \right)
      + \frac{\eta_0}{\rsun^2} \frac{\partial}{\partial\theta}\left[
        \frac{1}{\sin\theta}\frac{\partial}{\partial\theta}
        (\sin\theta\,b)\right] \,,
\label{eq:btor_1}
\end{eqnarray}
where $\Omega_{R_\odot}$ and $\Omega_{R_{\rm NSSL}}$ are the rotation rates as a function of latitude,
at the surface and at the bottom of the near-surface shear layer respectively; $V_\theta$ is the
effective velocity advecting the subsurface toroidal flux in the latitudinal direction; $\eta_0$
is the effective diffusivity of the azimuthal field in the latitudinal direction; and $\epsilon$
is a parameter which determines how much of the near-surface rotational shear affects the evolution
of the subsurface field.
The $\eta$ and $V_\theta$ appearing in this formula can include the effects of correlations between
the small scale field and flows, in particular turbulent diffusion and latitudinal pumping. The generation
of toroidal field by an alpha-effect is explicit assumed to be negigible.

Some of the quantities in these two formulae have been directly measured:
$U_{\theta,{R_\odot}}= -27.9 \cos\theta \sin\theta+ 17.7 \cos^3\theta \sin\theta$
\citep{2011ApJ...729...80H};
$\Omega_{R_{\odot}}=14.437-1.48\cos^2\theta-2.99\cos^4\theta$ \citep{2011ApJ...729...80H}; and
$\Omega_{R_{NSSL}}=14.18-1.59\cos^2\theta-2.61\cos^4\theta +0.53$, following
\cite{1998ApJ...505..390S} who give the surface rate and a near equator difference of $0.53^{\circ}$/day
between the surface and bottom of the near surface shear layer and \cite{2014A&A...570L..12B},
who report that the radial gradient at the top of the near-surface shear
layer is independent of latitude.

The surface diffusivity  $\eta_{R_\odot}$ due to the advection of the surface field by
the turbulent near-surface convective flows lies in the range of about 140~km$^2/$s
to 600~km$^2/$s \citep{Schrijver:Zwaan:2000}. In previous studies we have used a value of 250 km$^2$/s,
however for this paper we found a slightly better match with the observations for $\eta_{R_\odot}=500$~km$^2/$s.

The source of poloidal flux $a_s$ in the Babcock-Leighton framework is associated with Joy's law and
the emergence of active regions. We model it here as $a_s= \alpha_0 \sin\theta \cos\theta\,\, b(\theta,t)$.
For $V_{\theta}$ we choose the simplest functional form $V_{\theta}= V_0 \sin(2\theta)$.
The parameters, $\epsilon$, $\alpha_0$, and $V_0$, are not directly given by observations.
The effect of varying the parameters is discussed in \cite{2017A&A...599A..52C}. The solutions are not very
sensitive to the parameter $\epsilon$ which lies in the range of 0 to 1, and we set it
to 1 for this study (similar results can be found with $\epsilon=0$).

We choose $\alpha_0=1$~m/s, $V_0=2$~m/s and $\eta_0=90$~km$^2/$s because these values ($\alpha_0$, $\eta_0$, $V_0$) give a reasonable
correspondence of the model results to the observed azimuthally averaged radial and azimuthal field evolution,
as can be seen in Fig.~\ref{fig:sun}. The values used here seem plausible in comparison to the surface meridional
flow of about 15m/s, and the subsurface diffusivity of mixing length theory.

The linear Equations (2 -- 4) were solved numerically by discretizing the equations
using a five point stencil and looking for the fastest growing eigen-solutions. 
The match between the model and the observations, as shown in Fig.~\ref{fig:sun} is reasonable
in the sense that the model captures most of the evolution of the
large-scale field. Fine structure and cycle variability, introduced by the discreteness of emerge events,
is beyond the scope of the model.

Our conclusion for this modeling part of the paper is that the updated Babcock-Leighton
model presented in \cite{2017A&A...599A..52C} can explain important aspects of the observed evolution
of the poloidal and azimuthal fields (including the extended solar cycle). 

\begin{figure*}
\begin{center}
\includegraphics[scale=0.7]{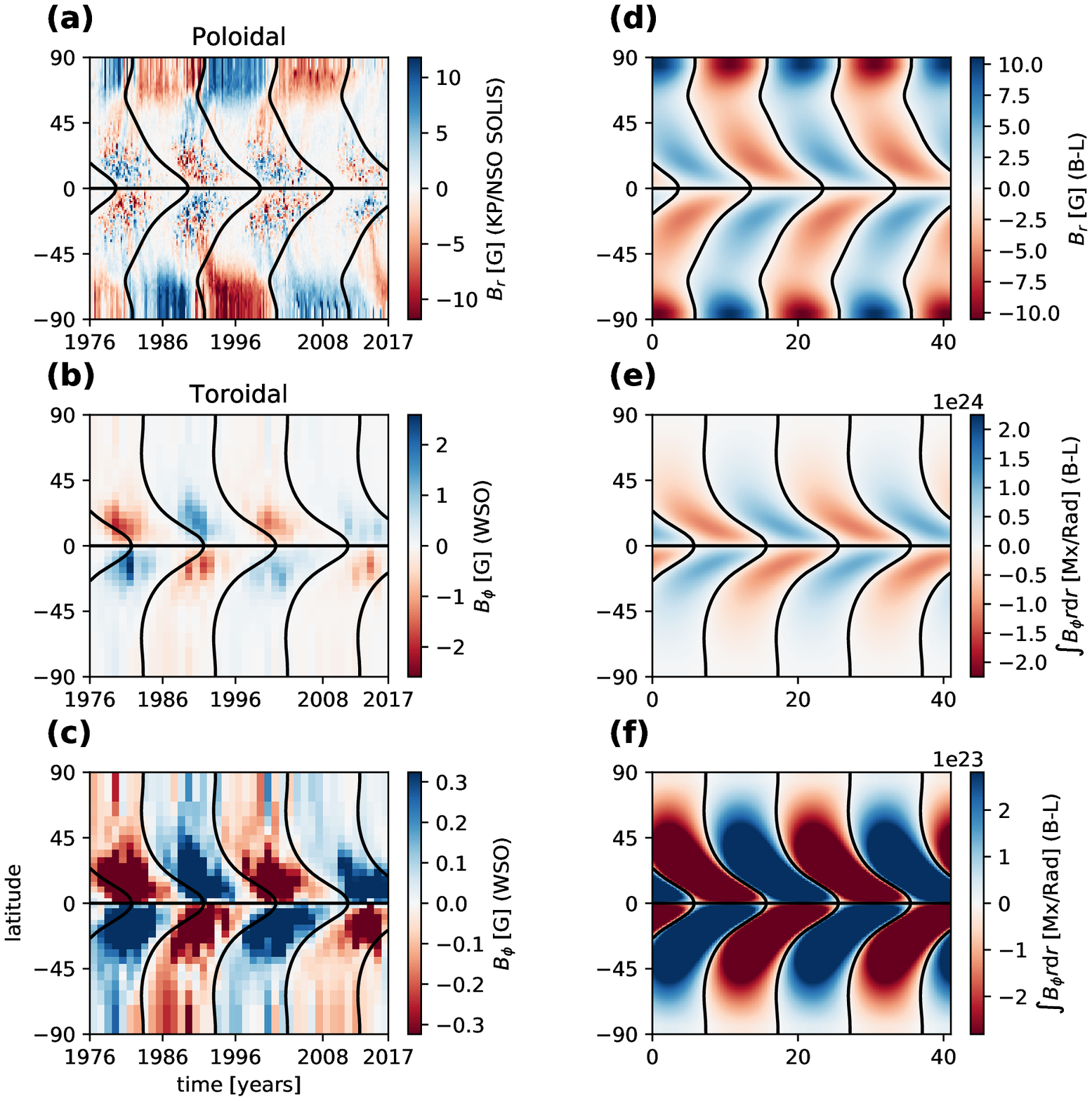}
\caption{ Similar to Fig.~\ref{fig:sun_decay} except that a simple decay of the poloidal and toroidal fields, with a decay time of
  22 years are included in the evolution equations for $a$ and $b$, and a diffusivity $\eta_0=38$~km$^{2}$s$^{-1}$ is assumed in the
  bulk of the convection zone.}
\label{fig:sun_decay}
\end{center}
\end{figure*}

An important remaining issue is that the transport of average longitudinally averaged azimuthal flux through the solar
photosphere during emergence reduces the subsurface toroidal field. To obtain an upper limit for the amount of flux lost,
we consider that all of the toroidal flux calculated above as passing through the solar photopsphere is lost to the interior.
The average toroidal field strength of 0.1~G (from the ephemeral regions) carried across the entire solar surface with an average
velocity of 1~km/s would carry 7.6$\times 10^{23}$~Mx of toroidal flux across the solar surface each 11 years. This should be
compared with 2.4$\times 10^{24}$~Mx of flux generated per (strong) cycle by the latitudinal differential rotation
\citep{2015Sci...347.1333C}. The loss of toroidal field through the photosphere is therefore at most modest.
In most flux transport dynamos, flux loss through the photosphere is included through a diffusive flux ( which is
consistent with the usually employed boundary condition where the toroidal field vanishes at the photosphere).
In the original Babcock-Leighton formulation of \cite{1969ApJ...156....1L}, this loss is treated with a simple decay term.
In the formulation of \cite{2017A&A...599A..52C},
used here, this loss term was ignored on the basis that the success of the surface flux transport model indicated that the
decay timescale should not be less than a few cycles \citep{2012A&A...542A.127C}, consistent with what we find here.
At such levels the decay term is not expected to produce qualitative changes in either the surface evolution of the radial field
\citep{2006A&A...446..307B} or the evolution of the subsurface toroidal field \citep{2015Sci...347.1333C}. To confirm this
Figure~\ref{fig:sun_decay} shows the result for a case where we have incuded a simple loss term
$-\frac{a}{\tau}$ and  $-\frac{b}{\tau}$ with $\tau=22$~years in Equations~3 and 5
and set $\eta_0=38$~km$^{2}$s$^{-1}$ so that the growth rate of the dynamo is positive
but close to zero (all other parameters were the same as for Fig.~\ref{fig:sun}). The results of this experiment are shown in
Fig.~\ref{fig:sun_decay}. Including this flux loss therefore does not qualitatively affect our conclusions.
    
\section{Conclusions}
We have almost 4 cycles of observations of both the radial and azimuthal magnetic fields relevant for
the global solar dynamo. To the extent that the solar dynamo
is a cycle where poloidal field is generated from toroidal field, followed by the generation of toroidal
field from poloidal field,  the observation of both these fields as a function of time and latitude, for almost
four solar cycles, is a stringent constraint for dynamo models. 

The (updated) Babcock-Leighton model is able to reproduce basic features of the observations with reasonable
values of the model's free parameters.

\begin{acknowledgements}
Data provided by the SOHO/MDI consortium. SOHO is a project of international cooperation between ESA and NASA.

Wilcox Solar Observatory data used in this study was obtained via the web site http://wso.stanford.edu, courtesy of J.T. Hoeksema.

NSO/Kitt Peak data used here are produced cooperatively
by NSF/NOAO, NASA/GSFC, and NOAA/SEL. This work utilizes SOLIS data
obtained by the NSO Integrated Synoptic Program (NISP), managed by the
National Solar Observatory, which is operated by the Association of
Universities for Research in Astronomy (AURA), Inc. under a cooperative
agreement with the National Science Foundation. 
\end{acknowledgements}
\bibliographystyle{aa}
\bibliography{TOR}

\begin{thebibliography}{31}
\expandafter\ifx\csname natexlab\endcsname\relax\def\natexlab#1{#1}\fi

\bibitem[{{Altrock}(1988)}]{1988sscd.conf..414A}
{Altrock}, R.~C. 1988, in Solar and Stellar Coronal Structure and Dynamics, ed.
  R.~C. {Altrock}, 414--420

\bibitem[{{Barekat} {et~al.}(2014){Barekat}, {Schou}, \&
  {Gizon}}]{2014A&A...570L..12B}
{Barekat}, A., {Schou}, J., \& {Gizon}, L. 2014, \aap, 570, L12

\bibitem[{{Baumann} {et~al.}(2006){Baumann}, {Schmitt}, \&
  {Sch{\"u}ssler}}]{2006A&A...446..307B}
{Baumann}, I., {Schmitt}, D., \& {Sch{\"u}ssler}, M. 2006, \aap, 446, 307

\bibitem[{{Cameron} \& {Sch{\"u}ssler}(2015)}]{2015Sci...347.1333C}
{Cameron}, R. \& {Sch{\"u}ssler}, M. 2015, Science, 347, 1333

\bibitem[{{Cameron} {et~al.}(2012){Cameron}, {Schmitt}, {Jiang}, \& {I{\c
  s}{\i}k}}]{2012A&A...542A.127C}
{Cameron}, R.~H., {Schmitt}, D., {Jiang}, J., \& {I{\c s}{\i}k}, E. 2012, \aap,
  542, A127

\bibitem[{{Cameron} \& {Sch{\"u}ssler}(2016)}]{2016A&A...591A..46C}
{Cameron}, R.~H. \& {Sch{\"u}ssler}, M. 2016, \aap, 591, A46

\bibitem[{{Cameron} \& {Sch{\"u}ssler}(2017)}]{2017A&A...599A..52C}
---. 2017, \aap, 599, A52

\bibitem[{{Carrington}(1858)}]{1858MNRAS..19....1C}
{Carrington}, R.~C. 1858, \mnras, 19, 1

\bibitem[{{Centeno}(2012)}]{2012ApJ...759...72C}
{Centeno}, R. 2012, \apj, 759, 72

\bibitem[{{Duvall}(1978)}]{1978PhDT.........4D}
{Duvall}, Jr., T.~L. 1978, PhD thesis, Stanford Univ., CA.

\bibitem[{{Duvall} {et~al.}(1979){Duvall}, {Scherrer}, {Svalgaard}, \&
  {Wilcox}}]{1979SoPh...61..233D}
{Duvall}, Jr., T.~L., {Scherrer}, P.~H., {Svalgaard}, L., \& {Wilcox}, J.~M.
  1979, \solphys, 61, 233

\bibitem[{{Guglielmino} {et~al.}(2012){Guglielmino}, {Mart{\'{\i}}nez Pillet},
  {Bonet}, {del Toro Iniesta}, {Bellot Rubio}, {Solanki}, {Schmidt},
  {Gandorfer}, {Barthol}, \& {Kn{\"o}lker}}]{2012ApJ...745..160G}
{Guglielmino}, S.~L., {Mart{\'{\i}}nez Pillet}, V., {Bonet}, J.~A., {et~al.}
  2012, \apj, 745, 160

\bibitem[{{Hagenaar}(2001)}]{2001ApJ...555..448H}
{Hagenaar}, H.~J. 2001, \apj, 555, 448

\bibitem[{{Hale} {et~al.}(1919){Hale}, {Ellerman}, {Nicholson}, \&
  {Joy}}]{1919ApJ....49..153H}
{Hale}, G.~E., {Ellerman}, F., {Nicholson}, S.~B., \& {Joy}, A.~H. 1919, \apj,
  49, 153

\bibitem[{{Harvey}(1992)}]{1992ASPC...27..335H}
{Harvey}, K.~L. 1992, in Astronomical Society of the Pacific Conference Series,
  Vol.~27, The Solar Cycle, ed. K.~L. {Harvey}, 335

\bibitem[{Hathaway(2015)}]{Hathaway2015}
Hathaway, D.~H. 2015, Living Reviews in Solar Physics, 12, 4

\bibitem[{{Hathaway} \& {Rightmire}(2011)}]{2011ApJ...729...80H}
{Hathaway}, D.~H. \& {Rightmire}, L. 2011, \apj, 729, 80

\bibitem[{{Howard}(1974)}]{1974SoPh...39..275H}
{Howard}, R. 1974, \solphys, 39, 275

\bibitem[{{Jiang} {et~al.}(2014){Jiang}, {Hathaway}, {Cameron}, {Solanki},
  {Gizon}, \& {Upton}}]{2014SSRv..186..491J}
{Jiang}, J., {Hathaway}, D.~H., {Cameron}, R.~H., {et~al.} 2014, \ssr, 186, 491

\bibitem[{{Leighton}(1969)}]{1969ApJ...156....1L}
{Leighton}, R.~B. 1969, \apj, 156, 1

\bibitem[{{Lo} {et~al.}(2010){Lo}, {Hoeksema}, \&
  {Scherrer}}]{2010ASPC..428..109L}
{Lo}, L., {Hoeksema}, J.~T., \& {Scherrer}, P.~H. 2010, in Astronomical Society
  of the Pacific Conference Series, Vol. 428, SOHO-23: Understanding a Peculiar
  Solar Minimum, ed. S.~R. {Cranmer}, J.~T. {Hoeksema}, \& J.~L. {Kohl}, 109

\bibitem[{{Martin} \& {Harvey}(1979)}]{1979SoPh...64...93M}
{Martin}, S.~F. \& {Harvey}, K.~H. 1979, \solphys, 64, 93

\bibitem[{{McIntosh} {et~al.}(2014){McIntosh}, {Wang}, {Leamon}, {Davey},
  {Howe}, {Krista}, {Malanushenko}, {Markel}, {Cirtain}, {Gurman}, {Pesnell},
  \& {Thompson}}]{2014ApJ...792...12M}
{McIntosh}, S.~W., {Wang}, X., {Leamon}, R.~J., {et~al.} 2014, \apj, 792, 12

\bibitem[{{Schou} {et~al.}(1998){Schou}, {Antia}, {Basu}, {Bogart}, {Bush},
  {Chitre}, {Christensen-Dalsgaard}, {Di Mauro}, {Dziembowski}, {Eff-Darwich},
  {Gough}, {Haber}, {Hoeksema}, {Howe}, {Korzennik}, {Kosovichev}, {Larsen},
  {Pijpers}, {Scherrer}, {Sekii}, {Tarbell}, {Title}, {Thompson}, \&
  {Toomre}}]{1998ApJ...505..390S}
{Schou}, J., {Antia}, H.~M., {Basu}, S., {et~al.} 1998, \apj, 505, 390

\bibitem[{{Schrijver} \& {Harvey}(1994)}]{1994SoPh..150....1S}
{Schrijver}, C.~J. \& {Harvey}, K.~L. 1994, \solphys, 150, 1

\bibitem[{{Schrijver} \& {Zwaan}(2000)}]{Schrijver:Zwaan:2000}
{Schrijver}, C.~J. \& {Zwaan}, C. 2000, {Solar and stellar magnetic activity}
  (Cambridge University Press)

\bibitem[{{Schwabe}(1849)}]{1849AN.....28..302S}
{Schwabe}, M. 1849, Astronomische Nachrichten, 28, 302

\bibitem[{{Shrauner} \& {Scherrer}(1994)}]{1994SoPh..153..131S}
{Shrauner}, J.~A. \& {Scherrer}, P.~H. 1994, \solphys, 153, 131

\bibitem[{{Sp{\"o}rer}(1879)}]{1879AN.....96...23S}
{Sp{\"o}rer}, F.~W.~G. 1879, Astronomische Nachrichten, 96, 23

\bibitem[{{Ulrich} \& {Boyden}(2005)}]{2005ApJ...620L.123U}
{Ulrich}, R.~K. \& {Boyden}, J.~E. 2005, \apjl, 620, L123

\bibitem[{{Wilson} {et~al.}(1988){Wilson}, {Altrocki}, {Harvey}, {Martin}, \&
  {Snodgrass}}]{1988Natur.333..748W}
{Wilson}, P.~R., {Altrocki}, R.~C., {Harvey}, K.~L., {Martin}, S.~F., \&
  {Snodgrass}, H.~B. 1988, \nat, 333, 748

\end{thebibliography}

\end{document}